\documentstyle[aps,preprint,eqsecnum,epsf]{revtex}
\tightenlines
\begin{document}
\draft
\title{Decuplet baryon magnetic moments\\ in a QCD-based quark model \\
beyond quenched approximation}
\author{
      Phuoc Ha\thanks{Electronic address: phuoc@theory1.physics.wisc.edu}}
\address{
        Physics Department, University of Wisconsin-Madison \\
             Madison, Wisconsin 53706, USA
        }
\maketitle

\begin{abstract}
We study the decuplet baryon magnetic moments in a QCD-based quark model
beyond quenched approximation. Our approach for unquenching the theory is
based on the heavy baryon perturbation theory in which the axial couplings for
baryon - meson and the meson-meson-photon couplings from the chiral
perturbation theory are used together with the QM moment couplings. It also
involves the introduction of a form factor characterizing the structure
of baryons considered as composite particles. Using the parameters obtained
from fitting the octet baryon magnetic moments, we predict the decuplet
baryon magnetic moments. The $\Omega^-$ magnetic moment is found to be in
good agreement with experiment: $\mu_{\Omega^-}$ is predicted to be $-1.97
\mu_N$ compared to the experimental result of ($-$2.02 $\pm$ 0.05) $\mu_N$.
\end{abstract}


\section{Introduction}

The naive, nonrelativistic quark model (NQM), even though very simple in
its formalism, is qualitatively good in describing the magnetic moments of
the octet baryons. It fits the pattern and the general magnitude of the
octet baryon moments up to $0.1\mu_N$ (nuclear magnetons) in average. The
discrepancies between theoretical predictions and experimental data are
due to the hadrons having an internal structure with a dynamically
intricate properties that the NQM has not accounted for. Therefore, it is
desirable to build a dynamical theory for the NQM.

In fact, the NQM can be derived from QCD using the Wilson loop approach
\cite{DandH1}. By calculating the gauge invariant Green's function for a baryon
interacting with an electromagnetic field and using well-defined
approximations, such as the ``quenched'' approximation in which the internal
virtual quark pair loops are not allowed and the minimal area law, we have
derived the quark model for moments plus semi relativistic corrections
associated with the binding of the quarks in the baryon. A test of this
QCD-based QM by fitting the octet baryon moments showed that the theory failed
to give any substantial improvement in the QM moments. The problem was
identifed with the quenched approximation \cite{DandH1}.

To go beyond the quenched approximation, we have developed an loop expansion
approach for the QCD-based QM and studied the octet baryon moments using
our newly developed approach \cite{DandH3}. Our calculation is based on the
heavy baryon perturbation theory in which the chiral baryon-meson couplings and
the meson-meson-photon couplings from the chiral perturbation theory together
with the QM moment couplings are used. It also involves the introduction of a
single form factor characterizing the structure of the baryons considered as
composite particles. The form factor reflects soft wave function effects with
characteristic momenta at a scale $\lambda \sim 400$ MeV, well below the chiral
cutoff $\sim 1$ GeV. We chose the strong interaction coupling constants in the
chiral baryon-meson couplings to satisfy the SU(6) relations $F=2/3D, \, {\cal
C}=-2D$, and ${\cal H}=-3D$, with $D=0.75$ as would be expected for the $L=0$
QM states. Our theory is convergent and has only three free parameters, the
effective quark moments $\mu_u, \mu_s$, and the wave function parameter
$\lambda$. The last is constrained by theory and experiment.
In contrast the usual approaches to magnetic moments through ChPT
\cite{Daietal,Mei,Bos,DandH2} involve seven parameters in the description of
the octet moments at one loop. If these parameters are used in fitting the
seven measured octet moments, the effects of dynamical loop corrections appear
only in the prediction for the $\Sigma^0\Lambda$ transition moment, where they
are small \cite{DandH2}.

We found in \cite{DandH3} that combining the dynamical
corrections from the loop expansion with those associated with the binding of
quarks in baryon significantly improved the agreement between the theoretical
and experimental values of the baryon magnetic moments. The average deviation
from fitting the seven well-measured octet magnetic moments excluding the
transition moment $\mu_{\Lambda\Sigma^0}$, is $0.05 \mu_N$, a substantial
improvement on the QM. We concluded that the loop expansion is an effective way
of going beyond quenched approximation in the octet baryon magnetic moment
problem.

In this paper, we study the decuplet baryon magnetic moments using the same
method. Our way of evaluating the semi relativistic corrections
associated with the binding of quarks in the baryon and the choice of the
strong interaction coupling constants and the octet - decuplet mass difference
are the same for both octet and decuplet. We can therefore evaluate the
decuplet moments using the quark moments $\mu_u, \mu_s$, and the wave function
parameter $\lambda$ obtained in fitting the octet baryon moments, and predict
the decuplet moments. In particular, the decuplet moment
$\mu_{\Omega^-}$ is predicted to be $-1.97 \mu_N$ compared to the
experimental result of ($-$2.02 $\pm$ 0.05) $\mu_N$. The loop
corrections are again small in comparison to the leading terms, and the
contributions from the decuplet intermediate states are substantial in
comparison to those from the octet intermediate states for some baryons.

The paper is organized as follows. Section II  briefly describes loop
expansion approach. An expression of the decuplet baryon magnetic moments are
given in Sec. III, where some numerical results of calculating the decuplet
baryon moments are also presented. The conclusions are given in Sec. IV. All
the necessary formulae for the decuplet baryon moments can be found in the
appendices.

\section{Loop expansion approach}

Going beyond the quenched approximation in the QCD-based QM means that we have
to develop an approach for studying the meson loop effects in the QCD-based QM.
We also need to take the composite structure of the baryons into account. This
is already included in the calculation of the QCD binding effects, but must
also be included in the loop calculations. For that purpose, we introduce a
single form factor characterizing the structure of the $L=0$ baryons considered
as composite particles. We base our loop calculations on heavy baryon
perturbation theory (HBPT) and use, together with the QM moment couplings,
chiral couplings for the low momentum couplings of mesons to baryons . That is,
the couplings of the heavy baryon chiral perturbation theory (HBChPT) are used
where chiral baryon-meson couplings and the meson-meson-electromagnetic field
couplings are invoked, but the actual calculation of the loop graphs is
modified with respect to \cite{Daietal}, \cite{Mei}.

\subsection{Definition of couplings}

\subsubsection{Chiral couplings}

HBChPT, which has been used to study the hadronic processes
of momentum transfers much less than 1 GeV, is well described in
Refs. \cite{HBChPT}. Let us consider a heavy baryon interacting with a low
momentum meson. The velocity of the baryon is nearly unchanged when it
exchanges some small momentum with the meson. Then, a nearly on-shell
baryon with velocity $v^\mu$ has momentum

\begin{equation}
  p^{\mu} = m_B v^{\mu} + k^{\mu} \ ,
\end{equation}
where $m_B$ is the baryon mass, and $k . v \ll m_B$. The effective heavy baryon
theory is written in terms of baryon fields $B_v$ with definite velocity
$v^\mu$, which are related to the original baryon fields by \cite{HBChPT}

\begin{equation}
  B_v (x) = e^{i m_B {\not v} v^{\mu} x_{\mu}}B(x) \ .
\end{equation}
The new baryon fields obey a modified Dirac equation,

\begin{equation}
  i{\not \partial}B_v =0 \ .
\end{equation}
The chiral Lagrangian for baryon fields depends on the
pseudoscalar meson octet

\begin{equation}\label{eq:1}
\bbox{\phi} = {1 \over \sqrt{2}} \left(
\begin{array}{ccc}
{\pi^0\over\sqrt{2}} + {\eta\over \sqrt{6}} & \pi^+ & K^+ \\ \pi^- & -{\pi^0
\over \sqrt{2}} + {\eta \over \sqrt{6}} & K^0 \\ K^- & \overline K^0 & -
{2\eta \over \sqrt{6}}
\end{array} \right) \ ,
\end{equation}
which couples to the baryon fields through the vector and axial vector
currents defined by

\begin{eqnarray}
V_\mu = {1 \over f^2} (\phi\partial_\mu\phi - \partial_\mu\phi \phi) + . . . \
, \ \
A_\mu&=&{ {\partial_\mu \phi} \over f} + ...  \ , \ \
\end{eqnarray}
where $f \sim 93$ MeV is the meson decay constant. We will retain, as shown
above, only leading term in the derivative expansion. The lowest order chiral
Lagrangian for octet and decuplet baryons is then
\begin{eqnarray}\label{lag}
{\cal L}_v &=& i \ {\rm Tr}\ \bar B_v\ \left(v\cdot {\cal D}
\right)B_v
+ 2\ D\  {\rm Tr}\ \bar B_v\ S_v^\mu\ \{ A_\mu, B_v \}
+ 2\ F\ {\rm Tr}\  \bar B_v\ S_v^\mu\ [A_\mu, B_v]
\nonumber \\
&&-\ i\ \bar
T_v^{\mu}\ (v \cdot {\cal D}) \  T_{v \mu}
+ \delta \ \bar T_v^{\mu}\ T_{v \mu}
+ {\cal C}\ \left(\bar T_v^{\mu}\ A_{\mu}\ B_v + \bar B_v\ A_{\mu}\
T_v^{\mu}\right){\phantom {f^2 \over 4}}
\nonumber\\
&& +\ 2\ {\cal H}\  \bar T_v^{\mu}\ S_{v \nu}\ A^{\nu}\  T_{v \mu}
+ {\rm Tr}\ \partial_\mu \phi \partial^\mu
\phi +\ \cdots \ , \ \
\end{eqnarray}
where $\delta$ is the decuplet-octet mass difference, and
${\cal D_\mu}= \partial_\mu+[V_\mu,  \; ]$ is the covariant chiral derivative.
$B_v$ is the usual matrix of octet baryons, and the $T_v^\mu$ are the
decuplet of baryons. $D$, $F$, ${\cal C}$, and ${\cal H}$ are the strong
interaction coupling constants. The spin operator $S_v^{\mu}$ is
defined in Ref.\cite{Jenetal}. This Lagrangian defines meson-baryon couplings
we will use.

The meson-meson-electromagnetic field couplings and the convection current
interactions of the baryons are introduced into the Lagrangian by making a
substitutions:
\begin{eqnarray}
{\cal D_\mu} &\rightarrow & {\cal D_\mu} + ie{\cal A}_\mu [ Q, ] \ , \nonumber
\\
\partial_\mu \phi & \rightarrow & {\cal D_\mu} \phi
         = \partial_\mu \phi + ie{\cal A}_\mu [ Q, \phi] \ , \ \
\end{eqnarray}
where ${\cal A}_\mu$ is the photon field.

\subsubsection{QM moment couplings}

In order to employ the techniques of HBPT, we need octet, decuplet, and
decuplet-octet transition magnetic moment operators which give the
corresponding QM moments. We can construct these using $B_v$, $T_v^\mu$,
and the moment operator $\widehat Q=$ diag $(\mu_u, \mu_d, \mu_s)$
\cite{DandH3}. For example, the QM decuplet magnetic moment operator is

\begin{equation} \label{decupmm}
{\cal L}^{(3/2)} = -i{3 e \over 2 m_N}
       \bar T_{v ikl}^{\mu} {\widehat Q}^i_j T_{v }^{\nu jkl} F^{\mu \nu} \ ,
\end{equation}
where $i$, $j$, $k$, and $l$ are SU(3) flavor indices. In a momentum space,
after doing a calculation on the flavor indices, we find that this operator
reproduces the QM decuplet moments
\begin{equation}
{\cal L}^{(3/2)}_b(q) = \mu_b^{QM} I \ ,
\end{equation}
where $q$ is the photon momentum and the spin dependent factor $I$ is defined
by
\begin{equation} \label{ftor}
  I =i\mu_N (\bar T^{'} \cdot {\cal A} T^{'} \cdot q -
\bar T^{'} \cdot q T^{'} \cdot {\cal A}) \ .
\end{equation}
The $T^{'}$'s are defined and the factor is evaluated in Appendix A using the
heavy baryon spin structure states. Note that the decuplet $T^{'}$'s are now
having the Dirac, spin and Lorentz indices only, $T^{'}=T^{'\mu}_{\alpha,
\lambda}$. The Dirac index $\alpha$ and spin index are suppressed.
The QM decuplet moments $\mu_b^{QM}$ are
\begin{eqnarray}
&&\mu_{\Delta^{++}}^{QM} = 3 \mu_u ,\quad
\mu_{\Delta^+}^{QM} = 2 \mu_u + \mu_d ,\quad
\mu_{\Delta^0}^{QM} = 2 \mu_d + \mu_u ,\quad
\mu_{\Delta^-}^{QM} = 3 \mu_d \ , \nonumber\\
&&\mu_{\Sigma^{*+}}^{QM} = 2 \mu_u + \mu_s  ,\quad
\mu_{\Sigma^{*0}}^{QM} = \mu_u + \mu_d + \mu_s ,\quad
\mu_{\Sigma^{*-}}^{QM} = 2 \mu_d + \mu_s \ , \nonumber\\
&&\mu_{\Xi^{*0}}^{QM} = 2 \mu_s + \mu_u ,\quad
\mu_{\Xi^{*-}}^{QM} = 2 \mu_s + \mu_d \ , \\
&&\mu_{\Omega^-}^{QM} = 3\mu_s. \nonumber
\end{eqnarray}
The decuplet-octet transition magnetic moment operator is chosen as

\begin{equation}
{\cal L}^{(od)} = -i{2 e \over m_N} F^{\mu \nu}(\epsilon_{ijk}
     {\widehat Q}_l^i \bar B_{vm}^j S_v^\mu T_v^{\nu klm} + h.c ) \ ,
\end{equation}
which gives the decuplet-octet transition moments
\begin{eqnarray}
\mu_{\Delta^{+}p} & = &{2\sqrt{2} \over 3}( \mu_u-\mu_d) \ , \ \ \ \ \ \ \ \
\mu_{\Delta^0 n } = {2\sqrt{2} \over 3}( \mu_u-\mu_d) , \nonumber \\
\mu_{\Sigma^{*+}\Sigma^{+} }& = &{2\sqrt{2} \over 3}( \mu_s-\mu_u) \ , \ \ \ \
\
 \mu_{\Sigma^{*-}\Sigma^{-}} = {2\sqrt{2} \over 3}( \mu_d-\mu_s) \ ,
\nonumber\\
\mu_{\Sigma^{*0}\Sigma^{0}} & = &{\sqrt{2} \over 3}( \mu_u+\mu_u-2\mu_s) \ , \
 \mu_{\Sigma^{*0}\Lambda} = \sqrt{{2 \over 3}}( \mu_d-\mu_u) \ , \\
\mu_{\Xi^{*0}\Xi^{0}} & = &{2\sqrt{2} \over 3}( \mu_s-\mu_u) \ , \ \ \ \ \ \
 \mu_{\Xi^{*-}\Xi^{-}} = {2\sqrt{2} \over 3}( \mu_d-\mu_s) \ , \nonumber
\end{eqnarray}
that are the same as the QM results except for a change in sign of
$\mu_{\Sigma^{*0}\Lambda}$ and $\mu_{\Xi^{*0}\Xi^{0}}$. This difference comes
from a difference choice of the phases of the baryon fields, and does not
affect to the calculations of the loop corrections for the baryon magnetic
moments.

\subsection{ Meson Wave Function Effects - Form Factor}

For investigating the meson wave function effects on the baryon moments, we
introduce at each vertex with a meson line a form factor $F(k,v)$ defined in
the rest frame of the heavy baryon by

\begin{equation} \label{formf}
F(k,v) = { \lambda^2 \over {\lambda^2 + {\bf k}^2}} \ ,
\end{equation}
where $ k=(k_0, {\bf k})$ is the 4-momentum of meson and $\lambda$ is a
parameter characterizing a natural momentum scale for the wave function,
expected to be much below 1 GeV. The form factor defined as in Eq.
(\ref{formf}) is normalized at chiral limit when {\bf k} is set equal to zero.
With the introduction of this form factor, all the Feynman integrals give
finite contributions. We therefore have a convergent theory in which the
counterterms characteristic of loop calculations in ChPT are no longer
necessary.

Our method for evaluating the Feynman integrals from the loop graphs (Figs.
\ref{fig:sqloop} and \ref{fig:logloop}) with the form factors inserted is as
follows.

First, we rewrite the form factor (\ref{formf}) in terms of $k^\mu$ and
$v^\mu$ as
\begin{equation}\label{formfr}
{- \lambda^2  \over k^2 -(k \cdot v)^2 - \lambda^2} \ .
\end{equation}
Then, using the Feynman parametrization formula, we combine the factors in the
denominator for the loop graph into a general form
\begin{equation}\label{deno}
k^2 + \alpha (k \cdot v)^2 + (k \cdot V) + C \ ,
\end{equation}
where $\alpha$ and $C$ are parameters independent of
the integral variables $k$, and the vector $V$ is any combination of $v$ and
the photon momentum $q$. At this point, by changing variables to
\begin{equation}\label{transf}
k^{'} = k + \beta v (k \cdot v) \ ,
\end{equation}
and choosing $\beta = \pm \sqrt{1+\alpha}-1$, we can get rid of the
$(k \cdot v)^2$ term in the denominator.  Eq.(\ref{deno}) becomes
\begin{equation}\label{deno1}
k^{'2} +(k^{'} \cdot \tilde V) + C \ .
\end{equation}
where the vector $\tilde V$ is also any combination of $v$ and $q$.
The Feynman integrals with the intergrands containing the denominators of
this type are easily evaluated. Note that the Jacobian of the transformation of
variables (Eq.(\ref{transf})) is $1/\sqrt{1+\alpha}$.


\section{Decuplet baryon magnetic moments}

\subsection{Theoretical expressions}

The calculation of the loop graphs shown in Figs. \ref{fig:sqloop} and
\ref{fig:logloop} is straightforward. The main difficulty is in the calculation
of the ``group coefficients'' that arise from the products of couplings. These
algebraic calculations were done using Mathematica and checked with some group
coefficients given in \cite{Butetal}. The results are given in Appendix B. We
will only give the final expressions for the decuplet baryon magnetic moments.
In units of nuclear magnetons, an expression of baryon moments
is given by

\begin{equation}
\mu_b = \mu_b^{(0)} + \mu_b^{(\delta=0)} + \mu_b^{(\delta \neq 0)} \ ,
\end{equation}
where $\mu_b^{(0)}$ are the contributions from the lowest loop order. These
include the QM moments plus the corrections $\Delta \mu_b^{QM}$ from the
QCD-based QM
\footnote{ The explicit expressions of $\Delta \mu_b^{QM}$ are given in
\cite{DandH1,DandH3}}. The terms in $\mu_b^{(\delta=0)}$ are contributions from
the loop graphs which are independent of the decuplet-octet mass difference
$\delta=m_B^{decuplet}-m_B^{octet}$ (here intermediate baryon states are purely
decuplet), and the terms in $\mu_b^{(\delta \neq 0)}$ are contributions from
the loop graphs dependent to $\delta$ (here intermediate baryon states are
octet or octet and decuplet combined). We find

\begin{eqnarray}
\mu_b^{(0)}&=&\alpha_b + \Delta \mu_b^{QM} \ , \\
\mu_b^{(\delta=0)} &=& \sum_{X=\pi,K} { m_N \over {72\pi f^2}}
        {\lambda^4 \over (\lambda+m_X)^3} \beta_b^{(X)}  \, \nonumber \\
       &+& \sum_{X=\pi,K,\eta}{1 \over {16\pi^2 f^2}}
         (\gamma_b^{1(X)}-2\lambda_b^{(X)}\alpha_b )
 L_0( m_X, \lambda) \ , \label{mu32}
\end{eqnarray}
and
\begin{eqnarray} \label{mu12}
\mu_b^{(\delta \neq 0)} &= & \sum_{X=\pi,K}- {m_N \over 16\pi f^2} \,
      \widetilde F(m_X ,   -\delta, \lambda)\tilde \beta_b^{(X)}  \nonumber \\
     &+& \sum_{X=\pi,K,\eta}{1 \over 32\pi^2 f^2}
  \Big [ \ ( \tilde \gamma_b^{1(X)}
  -2 \tilde \lambda_b^{(X)}\alpha_b ) L_1(m_X , -\delta, \lambda) \ +
    \tilde \gamma_b^{2(X)} L_2(m_X , -\delta, \lambda) \Big ]  \ ,
\end{eqnarray}
where $\alpha_b=\mu_b^{QM} $ \ , and the group coefficients $\beta_b^{(X)}$,
$\tilde \beta_b^{(X)}$, $\lambda_b^{(X)}$, $\tilde \lambda_b^{(X)}$,
$\gamma_b^{1(X)}$, $\tilde \gamma_b^{1(X)}$, and $\tilde \gamma_b^{2(X)}$
are given in the appendix B.

Analytic expressions for $ L_0(m_X,\lambda)$,
$\widetilde F(m_X,\delta,\lambda)$, $ L_1(m_X ,\delta,\lambda)$, and
$L_2(m_X ,\delta,\lambda)$, which are the functions of the meson masses,
the decuplet-octet mass difference $\delta$, and the natural cutoff
$\lambda$, are given in \cite{DandH3}. It is straightforward
to get $\widetilde F(m_X,-\delta,\lambda)$, $ L_1(m_X ,-\delta,\lambda)$, and
$L_2 (m_X ,-\delta,\lambda)$ from these expressions given, and such an example
is shown in Appendix C. To have an idea which corrections come from which
loop graphs (Figs. \ref{fig:sqloop} and \ref{fig:logloop}), it is necessary to
know that $\beta_b^{(X)}$, $\tilde \beta_b^{(X)}$, $\gamma_b^{1(X)}$,
$\tilde \gamma_b^{1(X)}$, $\tilde \gamma_b^{2(X)}$,  $\lambda_b^{(X)}$,
and $\tilde \lambda_b^{(X)}$  are the group coefficients of the graphs
1a, 1b, 2a, 2b, 2c (or 2d), 2e and 2f, respectively.

\subsection{Numerical results}

Now we are ready to evaluate the decuplet baryon magnetic moments. As done in
the octet moment case, the corrections $\Delta \mu_b^{QM}$
from the QCD-based QM are calculated using the values of $\epsilon$'s and
$\Delta$'s given in \cite{DandH2}. Again, for the loop corrections, the
coupling constants $F$, $D$, ${\cal C}$, and ${\cal H}$ were chosen such that
$F+D=1.25 \approx \mid g_A/g_V \mid$ ( $g_A$ and $g_V$ are the axial vector and
vector coupling constants, respectively) and the $SU(6)$ relations between the
coupling constants $ F=2D/3, {\cal C}=-2D$, and  ${\cal H}=-3D$ are satisfied,
as expected for $L=0$ baryons. We also choose the decuplet-octet mass
difference $\delta=250$ MeV and $f_\pi=93$ MeV, $f_K=f_\eta=1.2 f_\pi$. The
remaining three parameters $\mu_u$,
$\mu_s$, and the natural cutoff $\lambda$ are given the values that
give the best fit in the octet moment case, namely
$\mu_u=2.803\mu_N$, $\mu_s=-0.656\mu_N$, and $\lambda=407$ MeV.

We give our calculated values for the decuplet baryon magnetic moments, and the
corresponding values from the NQM, in Table I and a detailed breakdown of the
contributions of the loop integrals to the magnetic moments in Table II. We
find that the {\it predicted} decuplet moment
$\mu_{\Omega^-}  =-1.97 \mu_N$ is in very good agreement with the experimental
result of ($-$2.02 $\pm$ 0.05) $\mu_N$, and the theoretical value of
$\mu_{\Delta^{++}}=5.69\mu_N$ falls within the experimental range
(from 3.7 to 7.5 in unit of nuclear magnetons)

As in the octet case, again we see that the loop contributions are
small in comparison to the tree level or QM terms, that the contributions from
the graphs involving the intermediate decuplet states (sum of the graphs
1a, 2a, 2c, 2d, and 2e) are substantial. For some baryons, those
contributions are even larger than those from the graphs involving only
the intermediate octet states.

\section{Conclusions}

In this paper, we have extended our earlier calculations of the octet baryon
moments in a QCD-based QM with loop corrections to include the decuplet baryon
magnetic moments. We have predicted the decuplet moments
using the input parameters obtained from studying the octet baryon
moments. We find that our predicted decuplet moment $\mu_{\Omega^-}$ is in very
good agreement with its experimental value.

Again, we have shown that our loop approach for baryon magnetic moments in
a QCD-based QM works. The loop corrections extend our QCD-based QM beyond the
quenched approximation. The resulting theory describes the baryon magnetic
moments much better than the NQM. It can fit the seven observed octet baryon
magnetic moments up to about 0.05$\mu_N$ in average magnitude, gives a result
for the $\Sigma^0\Lambda$ transition moment consistent with experiment, and
predicts $\mu_{\Omega^-}$ very well. We hope that the other
decuplet baryon moments predicted from our theory will be tested by the
future experimental data.

\section*{ACKNOWLEDGMENTS}
The author would like to thank Professor Loyal Durand for helpful discussions
and valuable supports. This work was supported in part by the U.S. Department
of Energy under Grant No. DE-FG02-95ER40896.
\newpage

\appendix
\section{ Heavy baryon spin structure}
\setcounter{equation}{0}
In a rest frame of a spin-${3 \over 2}$ baryon, the states $ \mid j, j_z >$
of this baryon are specified by a vector $ {\bf e}$ and a spin-${1 \over 2}$
spinor $\xi_m$, $m=-{1 \over 2}, {1 \over 2}$  as follows

\begin{eqnarray} \label{decst}
 \mid {3 \over 2}, {3 \over 2} > &=& {\bf e}_{+1} \xi_{1 \over 2} \ , \nonumber
\\
 \mid {3 \over 2}, {1 \over 2} > &=& {1 \over \sqrt{3}}{\bf e}_{+1} \xi_{-{1
\over 2}} + \sqrt{{2 \over 3}}{\bf e}_0 \, \xi_{1 \over 2} \ , \nonumber \\
\mid {3 \over 2}, -{1 \over 2} > &=& \sqrt{{2 \over 3}}{\bf e}_0 \, \xi_{-{1
\over 2}} + {1 \over \sqrt{3}}{\bf e}_{-1} \xi_{1 \over 2} \ , \\
\mid {3 \over 2}, -{3 \over 2} > &=& {\bf e}_{-1} \xi_{-{1 \over 2}} \ .
\nonumber
\end{eqnarray}
These states are satisfied the expected orthogonality and normalization
properties. In terms of the vector-spinor functions $T^{'}=T^{'\mu}_{\alpha,
\lambda}$ with $\alpha$ a Dirac spinor index and $\lambda=j_z$ a total spin
index, the state $\mid {3 \over 2}, {3 \over 2} >$ is identified as
\begin{equation}
\mid {3 \over 2}, {3 \over 2} > =
T^{'+1}_{{1 \over 2},{3 \over 2}}=-{1 \over \sqrt{2}}(T^{'x}_{{1 \over 2},{3
\over 2}}+iT^{'y}_{{1 \over 2},{3 \over 2}}) \ ,
\end{equation}
and so on.

Consider the factor $ I = i\mu_N(\bar T^{'} \cdot {\cal A} T^{'} \cdot q - \bar
T^{'} \cdot q T^{'} \cdot {\cal A})$ that appears in Eq. (\ref{ftor}). In the
baryon rest frame $T^{'\mu} = (0, {\bf T}^{'})$, while ${\cal A}= (0, {\bf A})$
for a pure magnetic field, then the factor $I$ reduces to form
\begin{eqnarray} \label{factor}
  I &=& i\mu_N({\bf T}^{'*} \cdot {\bf A} {\bf T}^{'} \cdot {\bf q }- {\bf
T}^{'*} \cdot {\bf q } {\bf T }^{'}\cdot {\bf A}) \ , \nonumber \\
     &=& i\mu_N ( {\bf T}^{'*} \times {\bf T}^{'}) \cdot ({\bf A} \times {\bf
q}) \ ,\nonumber \\
     &=& \mu_N( {\bf T}^{'*} \times {\bf T}^{'}) \cdot {\bf B} \ ,
\end{eqnarray}
where ${\bf B}=i ({\bf A} \times {\bf q})$ is the magnetic field. By choosing
the magnetic field along the ${\bf e}_0$ direction, ${\bf B}= {\bf e}_0 B$,
then it follows from (\ref{decst}) and (\ref{factor})
\begin{eqnarray}
  I &=& \pm i\mu_N B \ \ \ {\rm for} \ \ \ j_z=\pm {3 \over 2} \ , \nonumber \\
     &=& \pm i \mu_N {B \over 3} \ \ \ {\rm for} \ \ \ j_z=\pm {1 \over 2} \ .
\end{eqnarray}

Using (\ref{decst}), we can check the validity of the following relation which
is useful when evaluating some loop graphs
\begin{equation}
  {\bar T }^{'\mu} \, [ q \cdot S_v, {\cal A} \cdot S_v] \, T^{'}_\mu = {1
\over 2} ({\bar T^{'}} \cdot {\cal A} \, T^{'} \cdot q -  {\bar T^{'}} \cdot q
\, T^{'}\cdot {\cal A}) \ ,
\end{equation}
where $S_v$ is the spin operator.

\section{The group coefficients}

In this appendix, the group coefficients are presented explicitly. For
simplicity, the superscript $(X)$ is suppressed. The group coefficients
$\beta_b$ evaluated from the graphs 1a, up to a factor ${\cal H}^2$, are

\begin{eqnarray} \label{betap}
 \beta_{\Delta^{++}}^{(\pi)} &=& {1 \over 3} \ , \ \ \ \
 \beta_{\Delta^+}^{(\pi)} = {1 \over 9} \ , \ \ \ \
 \beta_{\Delta^0}^{(\pi)} = - {1 \over 9} \ , \ \ \ \
 \beta_{\Delta^-}^{(\pi)} = -{1 \over 3} \ , \nonumber \\
 \beta_{\Sigma^{*+}}^{(\pi)} &=& {2 \over 9} \ , \ \
 \beta_{\Sigma^{*0}}^{(\pi)} = 0  \ , \ \
 \beta_{\Sigma^{*-}}^{(\pi)} = -{2 \over 9} \ , \nonumber \\
 \beta_{\Xi^{*0}}^{(\pi)} &=& {1 \over 9} \ , \ \ \ \
 \beta_{\Xi^{*-}}^{(\pi)} = -{1 \over 9}   \ , \\
 \beta_{\Omega^-}^{(\pi)} &=& 0 \ , \nonumber
\end{eqnarray}
for the pion loops,
\begin{eqnarray} \label{betak}
 \beta_{\Delta^{++}}^{(K)} &=& {1 \over 3} \ , \ \ \ \
 \beta_{\Delta^+}^{(K)} = {2 \over 9} \ , \ \ \ \
 \beta_{\Delta^0}^{(K)} = {1 \over 9} \ , \ \ \ \
 \beta_{\Delta^-}^{(K)} = 0 \ , \nonumber \\
 \beta_{\Sigma^{*+}}^{(K)} &=& {1 \over 9} \ , \ \
 \beta_{\Sigma^{*0}}^{(K)} = 0  \ , \ \
 \beta_{\Sigma^{*-}}^{(K)} = -{1 \over 9} \ , \nonumber \\
 \beta_{\Xi^{*0}}^{(K)} &=& -{1 \over 9} \ , \ \ \ \
 \beta_{\Xi^{*-}}^{(K)} = -{2 \over 9}   \ , \\
 \beta_{\Omega^-}^{(K)} &=& -{1 \over 3} \ , \nonumber
\end{eqnarray}
for the kaon loops. The group coefficients $\tilde \beta_b$ evaluated from
the graphs 1b, up to a factor ${\cal C}^2$, are

\begin{eqnarray} \label{tbetap}
 \tilde \beta_{\Delta^{++}}^{(\pi)} &=& 1 \ , \ \ \ \
 \tilde \beta_{\Delta^+}^{(\pi)} = {1 \over 3} \ , \ \ \ \
 \tilde \beta_{\Delta^0}^{(\pi)} = - {1 \over 3} \ , \ \ \ \
 \tilde \beta_{\Delta^-}^{(\pi)} = -1 \ , \nonumber \\
 \tilde \beta_{\Sigma^{*+}}^{(\pi)} &=& {2 \over 3} \ , \ \
 \tilde \beta_{\Sigma^{*0}}^{(\pi)} = 0  \ , \ \
 \tilde \beta_{\Sigma^{*-}}^{(\pi)} = -{2 \over 3} \ , \nonumber \\
 \tilde \beta_{\Xi^{*0}}^{(\pi)} &=& {1 \over 3} \ , \ \ \ \
 \tilde \beta_{\Xi^{*-}}^{(\pi)} = -{1 \over 3}   \ , \\
 \tilde \beta_{\Omega^-}^{(\pi)} &=& 0 \ , \nonumber
\end{eqnarray}
for the pion loops,
\begin{eqnarray} \label{tbetak}
 \tilde \beta_{\Delta^{++}}^{(K)} &=& 1 \ , \ \ \ \
 \tilde \beta_{\Delta^+}^{(K)} = {2 \over 3} \ , \ \ \ \
 \tilde \beta_{\Delta^0}^{(K)} = {1 \over 3} \ , \ \ \ \
 \tilde \beta_{\Delta^-}^{(K)} = 0 \ , \nonumber \\
 \tilde \beta_{\Sigma^{*+}}^{(K)} &=& {1 \over 3} \ , \ \
 \tilde \beta_{\Sigma^{*0}}^{(K)} = 0  \ , \ \
 \tilde \beta_{\Sigma^{*-}}^{(K)} = -{1 \over 3} \ , \nonumber \\
 \tilde \beta_{\Xi^{*0}}^{(K)} &=& -{1 \over 3} \ , \ \ \ \
 \tilde \beta_{\Xi^{*-}}^{(K)} = -{2 \over 3}   \ , \\
 \tilde \beta_{\Omega^-}^{(K)} &=& -1 \ , \nonumber
\end{eqnarray}
for the kaon loops. The group coefficients $\gamma_b^{1}$ evaluated from the
graphs 2a, up to a factor $11{\cal H}^2/9$, are

\begin{eqnarray} \label{gam1p}
 \gamma_{\Delta^{++}}^{1(\pi)} &=& 2 \mu_u \ , \ \ \ \
 \gamma_{\Delta^+}^{1(\pi)} = {13 \over 12} \mu_u \ , \ \ \ \
 \gamma_{\Delta^0}^{1(\pi)} = {\mu_u \over 6} \ , \ \ \ \
 \gamma_{\Delta^-}^{1(\pi)} = -{3 \over 4}\mu_u  \ , \nonumber \\
 \gamma_{\Sigma^{*+}}^{1(\pi)} &=& {1 \over 9} (5\mu_u + 4\mu_s)  \ , \ \
 \gamma_{\Sigma^{*0}}^{1(\pi)} = {2 \over 9} (\mu_u + 2\mu_s)  \ , \ \
 \gamma_{\Sigma^{*-}}^{1(\pi)} = {1 \over 9}(-\mu_u + 4 \mu_s )\ , \nonumber \\
 \gamma_{\Xi^{*0}}^{1(\pi)} &=& {\mu_s \over 3} \ , \ \ \ \
 \gamma_{\Xi^{*-}}^{1(\pi)} = {1 \over 12} (\mu_u +4 \mu_s)  \ , \\
 \gamma_{\Omega^-}^{1(\pi)} &=& 0 \ , \nonumber
\end{eqnarray}
for the pion loops,
\begin{eqnarray} \label{gam1k}
 \gamma_{\Delta^{++}}^{1(K)} &=& {1 \over 3}(2 \mu_u +\mu_s) \ , \ \
 \gamma_{\Delta^+}^{1(K)} = {1 \over 3} (\mu_u+\mu_s) \ , \ \
 \gamma_{\Delta^0}^{1(K)} = {\mu_s \over 3} \ , \ \ \ \
 \gamma_{\Delta^-}^{1(K)} = {1 \over 3}(-\mu_u+\mu_s)  \ , \nonumber \\
 \gamma_{\Sigma^{*+}}^{1(K)} &=& {1 \over 18} (29\mu_u + 16\mu_s)  \ , \ \
 \gamma_{\Sigma^{*0}}^{1(K)} = {4 \over 9} (\mu_u + 2\mu_s)  \ , \ \
 \gamma_{\Sigma^{*-}}^{1(K)} = {1 \over 18}(-13\mu_u + 16 \mu_s )\ , \nonumber
\\
 \gamma_{\Xi^{*0}}^{1(K)} &=& \mu_u+{5\mu_s \over 3} \ , \ \
 \gamma_{\Xi^{*-}}^{1(K)} = {1 \over 3} (-\mu_u +5 \mu_s)  \ , \\
 \gamma_{\Omega^-}^{1(K)} &=& {1 \over 6}(\mu_u+8\mu_s)  \ , \nonumber
   \end{eqnarray}
for the kaon loops, and
\begin{eqnarray} \label{gam1e}
\gamma_{\Delta^{++}}^{1(\eta)} &=& {\mu_u \over 2} \ , \ \ \ \
 \gamma_{\Delta^+}^{1(\eta)} = {\mu_u \over 4}  \ , \ \ \ \
 \gamma_{\Delta^0}^{1(\eta)} = 0 \ , \ \ \ \
 \gamma_{\Delta^-}^{1(\eta)} = -{ \mu_u \over 4}  \ , \nonumber \\
 \gamma_{\Sigma^{*+}}^{1(\eta)} &=& 0  \ , \ \
 \gamma_{\Sigma^{*0}}^{1(\eta)} = 0  \ , \ \
 \gamma_{\Sigma^{*-}}^{1(\eta)} = 0 \ , \nonumber \\
 \gamma_{\Xi^{*0}}^{1(\eta)} &=& {1 \over 6}(\mu_u + 2\mu_s) \ , \ \ \ \
 \gamma_{\Xi^{*-}}^{1(\eta)} = {1 \over 12} (-\mu_u +4 \mu_s)  \ , \\
 \gamma_{\Omega^-}^{1(\eta)} &=& 2\mu_s \ , \nonumber
\end{eqnarray}
for the $\eta$ loops. The coefficients $\tilde \gamma_b^{1}$ evaluated from
the graphs 2b are given, up to a factor ${\cal C}^2$, as follows

\begin{eqnarray} \label{tgam1p}
\tilde \gamma_{\Delta^{++}}^{1(\pi)} &=& {3 \over 2}\mu_u \ , \ \ \ \
\tilde \gamma_{\Delta^+}^{1(\pi)} = {2 \over 3} \mu_u \ , \ \ \ \
\tilde \gamma_{\Delta^0}^{1(\pi)} = -{\mu_u \over 6} \ , \ \ \ \
\tilde \gamma_{\Delta^-}^{1(\pi)} = -\mu_u  \ , \nonumber \\
\tilde \gamma_{\Sigma^{*+}}^{1(\pi)} &=& {7 \over 18} (2\mu_u + \mu_s)  \ , \ \
\tilde \gamma_{\Sigma^{*0}}^{1(\pi)}={1 \over 18} (2\mu_u + 7\mu_s)  \ , \ \
\tilde \gamma_{\Sigma^{*-}}^{1(\pi)}={1 \over 18}(-10\mu_u + 7\mu_s )\ ,
\nonumber \\
\tilde \gamma_{\Xi^{*0}}^{1(\pi)} &=& {2 \over 3}\mu_s \ , \ \ \ \
\tilde \gamma_{\Xi^{*-}}^{1(\pi)} = {1 \over 12} (-\mu_u +8 \mu_s)  \ , \\
\tilde \gamma_{\Omega^-}^{1(\pi)} &=& 0 \ , \nonumber
\end{eqnarray}
for the pion loops,
\begin{eqnarray} \label{tgam1k}
\tilde \gamma_{\Delta^{++}}^{1(K)} &=& {1 \over 3}(4\mu_u-\mu_s) \ , \ \
\tilde \gamma_{\Delta^+}^{1(K)} = {1 \over 3}(2\mu_u-\mu_s) \ , \ \ \ \
\tilde \gamma_{\Delta^0}^{1(K)} = -{\mu_s \over 3} \ , \ \ \ \
\tilde \gamma_{\Delta^-}^{1(K)} = -{1 \over 3}(2\mu_u+\mu_s) \ , \nonumber \\
\tilde \gamma_{\Sigma^{*+}}^{1(K)} &=& {1 \over 18} (7\mu_u + 8\mu_s)  \ , \ \
\tilde \gamma_{\Sigma^{*0}}^{1(K)}={1 \over 18} (\mu_u + 8\mu_s)  \ , \ \
\tilde \gamma_{\Sigma^{*-}}^{1(K)}={1 \over 18}(-5\mu_u + 8\mu_s )\ , \nonumber
\\
\tilde \gamma_{\Xi^{*0}}^{1(K)} &=& \mu_u+{\mu_s \over 3} \ , \ \ \ \
\tilde \gamma_{\Xi^{*-}}^{1(K)} = {1 \over 3} (-2\mu_u + \mu_s)  \ , \\
\tilde \gamma_{\Omega^-}^{1(K)} &=& {1 \over 6} (-\mu_u + 16\mu_s) \ ,
\nonumber
\end{eqnarray}
for the kaon loops, and
\begin{eqnarray} \label{tgam1e}
\tilde \gamma_{\Delta^{++}}^{1(\eta)} &=& 0 \ , \ \ \ \
\tilde \gamma_{\Delta^+}^{1(\eta)} = 0 \ , \ \ \ \
\tilde \gamma_{\Delta^0}^{1(\eta)} = 0 \ , \ \ \ \
\tilde \gamma_{\Delta^-}^{1(\eta)} = 0  \ , \nonumber \\
\tilde \gamma_{\Sigma^{*+}}^{1(\eta)} &=& {1 \over 6} (4\mu_u - \mu_s)  \ , \ \
\tilde \gamma_{\Sigma^{*0}}^{1(\eta)}={1 \over 6} (\mu_u - \mu_s)  \ , \ \
\tilde \gamma_{\Sigma^{*-}}^{1(\eta)}=-{1 \over 6}(2\mu_u + \mu_s )\ ,
\nonumber \\
\tilde \gamma_{\Xi^{*0}}^{1(\eta)} &=& {1 \over 6}(-\mu_u+4\mu_s) \ , \ \ \ \
\tilde \gamma_{\Xi^{*-}}^{1(\eta)} = {1 \over 12} (\mu_u +8 \mu_s)  \ , \\
\tilde \gamma_{\Omega^-}^{1(\eta)} &=& 0 \ , \nonumber
\end{eqnarray}
for the $\eta$ loops. The coefficients $\tilde \gamma_b^{2}$ evaluated from
the graphs 2c (or 2d) are given, up to a factor $2{\cal CH}/3$, by

\begin{eqnarray} \label{tgam2p}
\tilde \gamma_{\Delta^{++}}^{2(\pi)} &=& 2\mu_u \ , \ \ \ \
\tilde \gamma_{\Delta^+}^{2(\pi)} = {2 \over 3} \mu_u \ , \ \ \ \
\tilde \gamma_{\Delta^0}^{2(\pi)} = -{2 \over 3}\mu_u \ , \ \ \ \
\tilde \gamma_{\Delta^-}^{2(\pi)} = -2\mu_u  \ , \nonumber \\
\tilde \gamma_{\Sigma^{*+}}^{2(\pi)} &=& {4 \over 9} (\mu_u + 2\mu_s)  \ , \ \
\tilde \gamma_{\Sigma^{*0}}^{2(\pi)}={2 \over 9} (-\mu_u + 4\mu_s)  \ , \ \
\tilde \gamma_{\Sigma^{*-}}^{2(\pi)}={8 \over 9}(-\mu_u + \mu_s )\ , \nonumber
\\
\tilde \gamma_{\Xi^{*0}}^{2(\pi)} &=& {2 \over 3}\mu_s \ , \ \ \ \
\tilde \gamma_{\Xi^{*-}}^{2(\pi)} = {1 \over 3}(-\mu_u +2 \mu_s)   \ , \\
\tilde \gamma_{\Omega^-}^{2(\pi)} &=& 0 \ , \nonumber
\end{eqnarray}
for the pion loops,
\begin{eqnarray} \label{tgam2k}
\tilde \gamma_{\Delta^{++}}^{2(K)} &=& {4 \over 3}(\mu_u-\mu_s) \ , \ \ \ \
\tilde \gamma_{\Delta^+}^{2(K)} = {2 \over 3} (\mu_u-2\mu_s) \ , \ \ \ \
\tilde \gamma_{\Delta^0}^{2(K)} = -{4 \over 3}\mu_s \ , \ \ \ \
\tilde \gamma_{\Delta^-}^{2(K)} = -{2 \over 3} (\mu_u+2\mu_s)  \ , \nonumber \\
\tilde \gamma_{\Sigma^{*+}}^{2(K)} &=& {2 \over 9} (7\mu_u - 4\mu_s)  \ , \ \
\tilde \gamma_{\Sigma^{*0}}^{2(K)}={2 \over 9} (\mu_u - 4\mu_s)  \ , \ \
\tilde \gamma_{\Sigma^{*-}}^{2(K)}=-{2 \over 9}(5\mu_u + 4\mu_s )\ , \nonumber
\\
\tilde \gamma_{\Xi^{*0}}^{2(K)} &=& {4 \over 3}\mu_s \ , \ \ \ \
\tilde \gamma_{\Xi^{*-}}^{2(K)} = {2 \over 3} (-\mu_u +2 \mu_s)  \ , \\
\tilde \gamma_{\Omega^-}^{2(K)} &=& {2 \over 3} (-\mu_u +4 \mu_s) \ , \nonumber
\end{eqnarray}
for the kaon loops, and
\begin{eqnarray} \label{tgam2e}
\tilde \gamma_{\Delta^{++}}^{2(\eta)} &=& 0 \ , \ \ \ \
\tilde \gamma_{\Delta^+}^{2(\eta)} = 0 \ , \ \ \ \
\tilde \gamma_{\Delta^0}^{2(\eta)} = 0 \ , \ \ \ \
\tilde \gamma_{\Delta^-}^{2(\eta)} = 0  \ , \nonumber \\
\tilde \gamma_{\Sigma^{*+}}^{2(\eta)} &=& 0  \ , \ \
\tilde \gamma_{\Sigma^{*0}}^{2(\eta)}=0  \ , \ \
\tilde \gamma_{\Sigma^{*-}}^{2(\eta)}=0 \ , \nonumber \\
\tilde \gamma_{\Xi^{*0}}^{2(\eta)} &=& {2 \over 3}(\mu_u-\mu_s) \ , \ \ \ \
\tilde \gamma_{\Xi^{*-}}^{2(\eta)} = -{1 \over 3} (\mu_u +2 \mu_s)  \ , \\
\tilde \gamma_{\Omega^-}^{2(\eta)} &=& 0 \ , \nonumber
\end{eqnarray}
for the $\eta$ loops. The group coefficients $\lambda_b$ evaluated from
the graphs 2e, up to a factor ${\cal H}^2$, are

\begin{equation} \label{lamp}
 \lambda_{\Delta}^{(\pi)} = {25 \over 36} \ ,
 \lambda_{\Sigma^{*}}^{(\pi)} = {10 \over 27} \ ,
 \lambda_{\Xi^{*}}^{(\pi)} = {5 \over 36} \ ,
 \lambda_{\Omega^-}^{(\pi)} = 0 \ ,
\end{equation}
for the pion loops
\begin{equation} \label{lamk}
 \lambda_{\Delta}^{(K)} = {5 \over 18} \ ,
 \lambda_{\Sigma^{*}}^{(K)} = {20 \over 27} \ ,
 \lambda_{\Xi^{*}}^{(K)} = {5 \over 6} \ ,
 \lambda_{\Omega^-}^{(K)} = {5 \over 9} \ ,
\end{equation}
for the kaon loops, and
\begin{equation} \label{lame}
 \lambda_{\Delta}^{(\eta)} = {5 \over 36} \ ,
 \lambda_{\Sigma^{*}}^{(\eta)} = 0  \ ,
 \lambda_{\Xi^{*}}^{(\eta)} = {5 \over 36} \ ,
 \lambda_{\Omega^-}^{(\eta)} = {5 \over 9} \ ,
\end{equation}
for the $\eta$ loops. The group coefficients $\tilde \lambda_b$ evaluated from
the graphs 2f, up to a factor ${\cal C}^2$, are

\begin{equation} \label{tlamp}
 \tilde \lambda_{\Delta}^{(\pi)} = {1 \over 2} \ ,
 \tilde \lambda_{\Sigma^{ \tilde*}}^{(\pi)} = {5 \over 12} \ ,
 \tilde \lambda_{\Xi^{*}}^{(\pi)} = {1 \over 4} \ ,
 \tilde \lambda_{\Omega^-}^{(\pi)} = 0 \ ,
\end{equation}
for the pion loops
\begin{equation} \label{tlamk}
 \tilde \lambda_{\Delta}^{(K)} = {1 \over 2} \ ,
 \tilde \lambda_{\Sigma^{*}}^{(K)} = {1 \over 3} \ ,
 \tilde \lambda_{\Xi^{*}}^{(K)} = {1 \over 2} \ ,
 \tilde \lambda_{\Omega^-}^{(K)} = 1 \ ,
\end{equation}
for the kaon loops, and
\begin{equation} \label{tlame}
 \tilde \lambda_{\Delta}^{(\eta)} = 0 \ ,
 \tilde \lambda_{\Sigma^{*}}^{(\eta)} = 1  \ ,
 \tilde \lambda_{\Xi^{*}}^{(\eta)} = {1 \over 4} \ ,
 \tilde \lambda_{\Omega^-}^{(\eta)} = 0 \ ,
\end{equation}
for the $\eta$ loops.
\newpage

\section{The expressions of $\widetilde F$, $ L_0$, $L_1$, and $L_2$}

The expressions of $ L_0(m,\lambda)$,
$\widetilde F(m,\delta,\lambda)$, $ L_1(m_X ,\delta,\lambda)$, and
$L_2(m_X ,\delta,\lambda)$ are given in \cite{DandH3}. In order to get
$\widetilde F(m_X,-\delta,\lambda)$, $ L_1(m_X ,-\delta,\lambda)$, and
$L_2 (m_X ,-\delta,\lambda)$ from them, we make an analytic continuation from
positive to negative $\delta$. Note that the functions can acquire an imaginary
part in the continuation, but it will not contribute to the decuplet moments
and therefore can be ignored. The real parts of the new functions are obtained
by a simple substitution of $-\delta$ for $\delta$.

As an illustration, The function $\widetilde F(m,-\delta,\lambda)$ is  found of
the form

\begin{eqnarray} \label{funf}
\pi \widetilde F(m,-\delta,\lambda) &=& {\lambda^4 \over
3(\lambda^2-m^2+\delta^2)^2}
   \left \{ -N(m,-\delta,\lambda)+{5\lambda^2+2m^2 \over \lambda^2-m^2}\delta +
    {\lambda^2+2m^2 \over (\lambda^2-m^2)^2}\delta^3 \right . \nonumber \\
  &+& {\lambda\delta \over (\lambda^2-m^2)^2(\lambda^2-m^2+\delta^2)}
     \Big [ 3(2\lambda^2+3m^2)(\lambda^2-m^2)
     -2(\lambda^2-6m^2)\delta^2 \nonumber \\
  &+& \left . {3 m^2 \over \lambda^2-m^2}\delta^4 \Big ] \,
     F_0 (m, \lambda) \right \} \ ,
\end{eqnarray}
where
\begin{equation}
N(m,-\delta,\lambda) = {1 \over (\lambda^2-m^2+\delta^2)}
             \Big [ \pi\lambda(\lambda^2+3m^2-3\delta^2)
            - 2(3\lambda^2+m^2-\delta^2) \, F_0 (m, -\delta) \Big ] \ ,
\end{equation}
and
\begin{eqnarray}
 F_0 (m, \pm \delta) & = &  \sqrt{m^2-\delta^2} \, ( \pi/2 \mp {\rm arctan} \,
          [\delta / \sqrt{m^2-\delta^2}]) \hskip 2.6cm {\rm for} \ \ \ m \geq
\delta  \ , \\
       & = &   \sqrt{\delta^2-m^2} \, \{ {\rm ln} \,
   [(\mp \delta +\sqrt{\delta^2-m^2}) / m]- (1 \mp 1) i\pi/2 \}  \hskip 0.5cm
{\rm for} \ \ \ m < \delta \ .
\end{eqnarray}

%

Similarly, the functions $ L_1(m,-\delta,\lambda)$, and
$L_2(m,-\delta,\lambda)$ can be easily read off from $ L_1(m,\delta,\lambda)$,
and
$L_2(m,\delta,\lambda)$.

\newpage

\begin{figure}
\caption{Diagrams that give rise to non-analytic
$m_s^{1/2}$ corrections to the baryon magnetic moments in the conventional
ChPT. The dashed lines denote the mesons, the single and double solid lines
denote octet and decuplet baryons, respectively. A heavy dot with a meson
line represents a form factor $F(k,v)$ (Eq.(\protect\ref{formf})), where k is
the meson momentum.}
\label{fig:sqloop}
\end{figure}

\begin{figure}
\caption{Diagrams that give rise to non-analytic
$m_s \ln m_s$ corrections to the baryon magnetic moments in the
conventional ChPT.}
\label{fig:logloop}
\end{figure}

\newpage

\centerline{
\epsffile{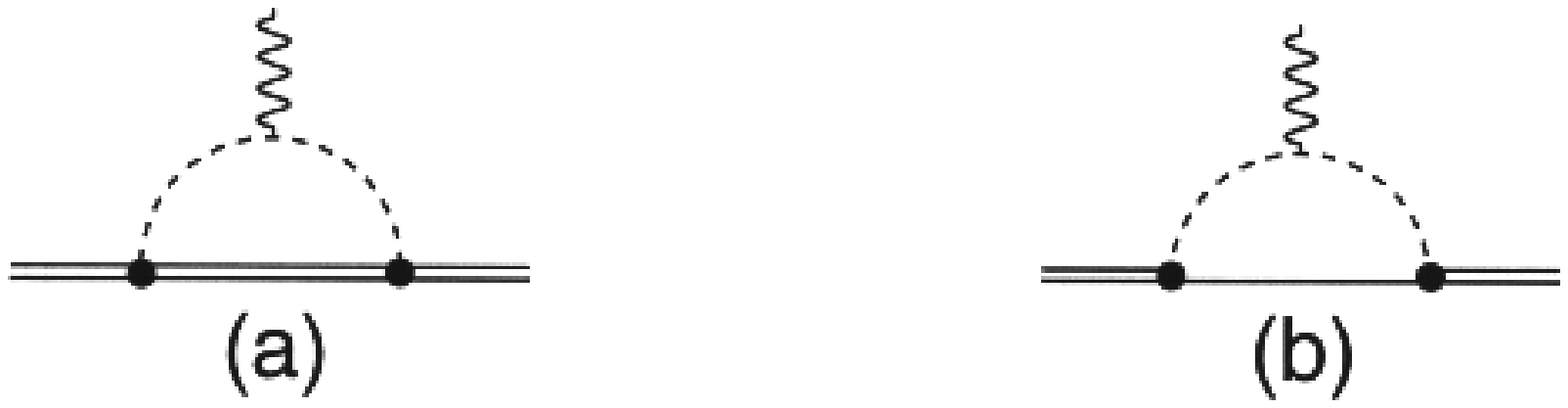}
}

\newpage

\centerline{
\epsffile{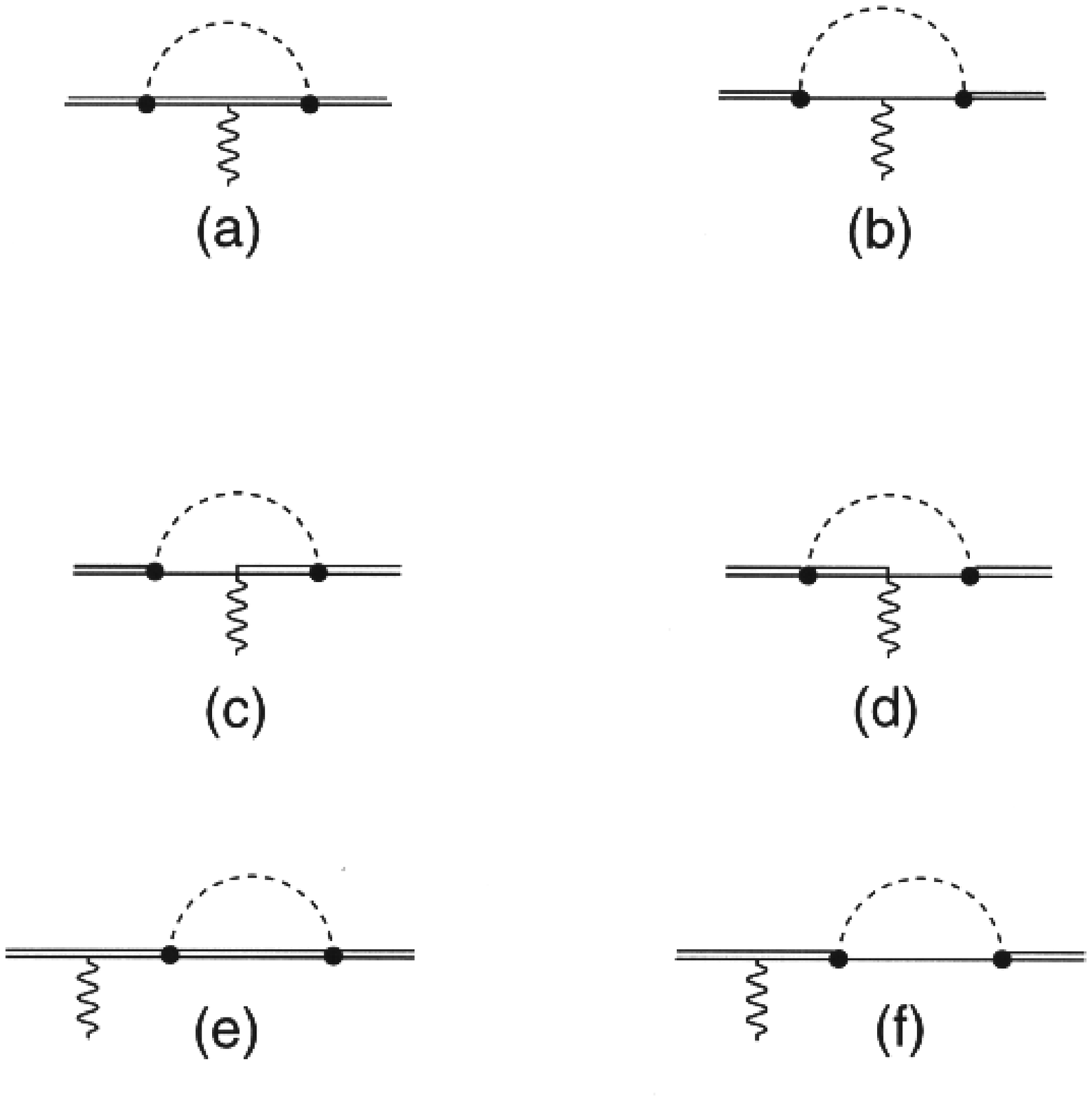}
}

\newpage
\medskip


\begin{table}
\caption{ The decuplet magnetic moments from the QM and the QCD-based QM with
loop corrections.}
\smallskip
\begin{center}
\begin{tabular}{crrr}
     $\mu_b$ & QM &  QM w/ loops & Experiment \\
    \hline
    $\Delta^{++}$     &  5.455   &  5.689  &  $3.5 - 7.5$ \\
    $\Delta^+$        &  2.728   &  2.778  &  $-$ \\
    $\Delta^0$        &  0 & $-$0.134   &  $-$ \\
    $\Delta^-$        &  $-$2.728   & $-$3.045  &  $-$ \\
    $\Sigma^{*+}$     &  3.057   &  2.933  &  $-$ \\
    $\Sigma^{*0}$     &  0.329   &  0.137  &  $-$ \\
    $\Sigma^{*-}$     &  $-$2.399   &  $-$2.659  &  $-$ \\
    $\Xi^{*0}$        &  0.658   &  0.424  &  $-$ \\
    $\Xi^{*-}$        &  $-$2.069   &  $-$2.307  &  $-$ \\
    $\Omega^-$        &  $-$1.740   &  $-$1.970  &
$-$2.02 $\pm$ 0.05  \\
     \end{tabular}
\end{center}
\end{table}
\medskip


\begin{table}

\caption{ Detailed breakdown of the contributions of the loop integrals
to the magnetic moments of the decuplet baryons (in $\mu_N$). Those
contributions are evaluated at $F=0.5$, $D=0.75$, $ {\cal C} =-1.5$, ${\cal H}
= -2.25$, $\delta=250 $ MeV, $\mu_u=2.083$, $\mu_s=-0.656$ and the natural
cutoff $\mu =407$ MeV.}
\smallskip
\begin{tabular}{lrrrrrrrr}
     $\mu_b$ & $\mu_u$, $\mu_s$ & $\Delta \mu_b^{QM}$ & $m_s^{1/2(N)}$ & ln
$m_s^{(N)}$ & $m_s^{1/2 (\Delta)}$ & ln $m_s^{(\Delta)}$ & Loops & $\mu_b$  \\
    \hline
    $\Delta^{++}$     &  6.249 & $-$0.434 & 0.078  & $-$0.351  &  0.159 &
$-$0.012 & $-$0.126 & 5.689 \\
    $\Delta^{+}$       & 3.125 & $-$0.217 & 0.052 & $-$0.183 & 0.060 & $-$0.059
& $-$0.130 & 2.778  \\
    $\Delta^{0}$  &  0 & 0 & 0.026 & $-$0.015 & $-$0.039 & $-$0.106 & $-$0.134
& $-$0.134  \\
    $\Delta^{-}$  & $-$3.125 & 0.217 &0 & 0.153 & $-$0.138 & $-$0.152 &
$-$0.138 & $-$3.045  \\
     $\Sigma^{*+}$ & 3.510 & $-$0.343 & 0.026  & $-$0.192 &0.099 & $-$0.167 &
$-$0.234 & 2.933  \\
    $\Sigma^{*0}$   & 0.386 & $-$0.089 &0 & $-$0.032 & 0 & $-$0.127 & $-$0.159
& 0.137 \\
    $\Sigma^{*-}$     & $-$2.739 & 0.165 &$-$0.026 & 0.127 & $-$0.099 &
$-$0.087 & $-$0.085 & $-$2.659  \\
    $\Xi^{*0}$ & 0.771 & $-$0.191 & $-$0.026 & $-$0.052 & 0.039 & $-$0.117 &
$-$0.156 & 0.424 \\
    $\Xi^{*-}$ & $-$2.354  & 0.096 & $-$0.052 & 0.102 & $-$0.060 & $-$0.041 &
$-$0.050 & $-$2.307  \\
    $\Omega^-$  & $-$1.968  & 0.013 & $-$0.077 & 0.077 & $-$0.021 & 0.006 &
$-$0.015 & $-$1.970

  \end{tabular}

\end{table}

\end{document}